\documentclass[]{article}
\usepackage{zeldovich}
\usepackage{amssymb}
\usepackage{bm}
\usepackage{epsf}
\usepackage{graphicx}
\usepackage{cite}

\def\be{\begin{equation}}
\def\ee{\end{equation}}
\def\bea{\begin{eqnarray}}
\def\eea{\end{eqnarray}}

\begin{document}

\title{SOLUTION OF THE DISCRETE WHEELER-DEWITT EQUATION IN THE
VICINITY OF SMALL SCALE FACTORS AND QUANTUM MECHANICS IN THE SPACE
OF NEGATIVE CONSTANT CURVATURE }

\author{S.L. Cherkas\\{\small\itshape
Institute for Nuclear Problems, Belarusian State University,}
\\{\small\itshape Bobruiskaya 11, Minsk 220050, Belarus}\\
V.L. Kalashnikov\\
{\small\itshape Institute for Photonics , Vienna University of
Technology,}\\{ \small\itshape Gusshausstrasse 27/387, Vienna
A-1040, Austria}}

\maketitle

\begin{abstract}{The asymptotic of the
solution of the discrete Wheeler-DeWitt equation is found in the
vicinity of small scale factors. It is shown that this problem is
equivalent to the solution of the stationary Schr\"{o}dinger
equation in the (super) space of negative constant curvature. The
minimum positive eigenvalue is found from which a continuous spectrum
begins. }\end{abstract}

\bigskip

The Wheeler-DeWitt equation  [1,2] is a functional equation describing
quantum space-time. As the solution of this equation
presents great difficulties, it is important to
investigate at least asymptotic of the solutions, for example, in
the vicinity of small scale factors. Under the scale factor it is
understood $a\left( {\bm r} \right) \equiv \gamma ^{1/6}\left(
{\bm r} \right)$, where $\gamma = \det\gamma _{ij} $. A metric
tensor $\gamma _{ij} \left( {\bm r} \right)$ depends on
coordinates $\bm r$, which are defined on the three-dimensional manifold.
The Wheeler-DeWitt equation for gravitation and several scalar fields
in the vicinity of small scale factors $a \sim 0$ looks as

\begin{equation}
\label{eq1} \left( {\frac{{12\gamma ^{1/6}}}{{M_{p}^{2}} }G_{ijkl}
\frac{{\delta }}{{\delta \gamma _{ij} \left( {\bm r}
\right)}}\frac{{\delta} }{{\delta \gamma _{kl} \left( {\bm r}
\right)}} + \frac{{\gamma ^{ - 1/3}}}{{2}}\frac{{\delta }}{{\delta
\phi \left( {\bm r} \right)}}\frac{{\delta} }{{\delta \phi \left(
{\bm r} \right)}}} \right)\Psi \left[ {\gamma ,\phi}  \right] = 0,
\end{equation}

\noindent where $\phi \left( {\bm r} \right) = \{ \phi _{1} \left( {\bm r}
\right),\phi _{2} \left( {\bm r} \right)...\phi _{N} \left( {\bm
r} \right)\} $ is a set of scalar fields, $M_{p} $ is the Planck mass
and
\[
G_{ijkl} = \frac{{1}}{{2}}\gamma ^{ - 1/2}\left( {\gamma _{ik}
\gamma _{jl} + \gamma _{il} \gamma _{jk} - \gamma _{ij} \gamma
_{kl}}  \right).
\]

Eq. (\ref{eq1}) is written in the conformal time gauge
[3]. As the Wheeler-DeWitt equation contains the functional
derivatives acting at the same spatial point, generally, a regularization is required to avoid
an occurrence of infinite quantities.
Besides it is necessary to choose an operator ordering procedure. The
most natural  choice of operator ordering is to form a
multivariate Laplacian. Let as write down all variables in the
form of a single vector $\bm \xi = \{ \gamma _{11} ,\gamma _{22}
,\gamma _{33} ,\gamma _{12} ,\gamma _{13} ,\gamma _{23} ,\phi _{1}
, \ldots \phi _{N} \} $. Then, Eq. (\ref{eq1}) takes the
form:
\begin{equation}
\label{eq2} G^{AB}\frac{{\delta} }{{\delta \xi ^{A}\left( {\bm r}
\right)}}\frac{{\delta }}{{\delta \xi ^{B}\left( {\bm r}
\right)}}\Psi \left[ {\bm \xi}(\bm r)  \right] = 0,
\end{equation}
where a matrix $G^{AB}$ is defined by
\begin{equation}
\label{eq3} G = \left( {{\begin{array}{*{20}c}
 {\mathop {\tilde {G}}\nolimits_{1111}}  \hfill & {\mathop {\tilde
{G}}\nolimits_{1122}}  \hfill & {\mathop {\tilde
{G}}\nolimits_{1133}} \hfill & {\mathop {\tilde
{G}}\nolimits_{1112}}  \hfill & {\mathop {\tilde
{G}}\nolimits_{1113}}  \hfill & {\mathop {\tilde
{G}}\nolimits_{1123}}
\hfill & {0} \hfill & {0} \hfill \\
 {\mathop {\tilde {G}}\nolimits_{2211}}  \hfill & {\mathop {\tilde
{G}}\nolimits_{2222}}  \hfill & {\mathop {\tilde
{G}}\nolimits_{2233}} \hfill & {\mathop {\tilde
{G}}\nolimits_{2212}}  \hfill & {\mathop {\tilde
{G}}\nolimits_{2213}}  \hfill & {\mathop {\tilde
{G}}\nolimits_{2223}}
\hfill & {0} \hfill & {0} \hfill \\
 {\mathop {\tilde {G}}\nolimits_{3311}}  \hfill & {\mathop {\tilde
{G}}\nolimits_{3322}}  \hfill & {\mathop {\tilde
{G}}\nolimits_{3333}} \hfill & {\mathop {\tilde
{G}}\nolimits_{3312}}  \hfill & {\mathop {\tilde
{G}}\nolimits_{3313}}  \hfill & {\mathop {\tilde
{G}}\nolimits_{3323}}
\hfill & {0} \hfill & {0} \hfill \\
 {\mathop {\tilde {G}}\nolimits_{1211}}  \hfill & {\mathop {\tilde
{G}}\nolimits_{1222}}  \hfill & {\mathop {\tilde
{G}}\nolimits_{1233}} \hfill & {\mathop {\tilde
{G}}\nolimits_{1212}}  \hfill & {\mathop {\tilde
{G}}\nolimits_{1213}}  \hfill & {\mathop {\tilde
{G}}\nolimits_{1223}}
\hfill & {0} \hfill & {0} \hfill \\
 {\mathop {\tilde {G}}\nolimits_{1311}}  \hfill & {\mathop {\tilde
{G}}\nolimits_{1322}}  \hfill & {\mathop {\tilde
{G}}\nolimits_{1333}} \hfill & {\mathop {\tilde
{G}}\nolimits_{1312}}  \hfill & {\mathop {\tilde
{G}}\nolimits_{1313}}  \hfill & {\mathop {\tilde
{G}}\nolimits_{1323}}
\hfill & {0} \hfill & {0} \hfill \\
 {\mathop {\tilde {G}}\nolimits_{2311}}  \hfill & {\mathop {\tilde
{G}}\nolimits_{2322}}  \hfill & {\mathop {\tilde
{G}}\nolimits_{2333}} \hfill & {\mathop {\tilde
{G}}\nolimits_{2312}}  \hfill & {\mathop {\tilde
{G}}\nolimits_{2313}}  \hfill & {\mathop {\tilde
{G}}\nolimits_{2323}}
\hfill & {0} \hfill & {0} \hfill \\
 {0} \hfill & {0} \hfill & {0} \hfill & {0} \hfill & {0} \hfill & {0} \hfill
& {\gamma ^{ - 1/3}} \hfill & {0} \hfill \\
 {0} \hfill & {0} \hfill & {0} \hfill & {0} \hfill & {0} \hfill & {0} \hfill
& {0} \hfill & {\gamma ^{ - 1/3}} \hfill \\
\end{array}} } \right),
\end{equation}
$\mathop {\tilde {G}}\nolimits_{ijkl} = {\textstyle{{12\gamma
^{1/6}} \over {M_{p}^{2}} }}G_{ijkl} ,$ and Eq.
(\ref{eq3}) is written down for a special case of two scalar
fields.

Eqs. (\ref{eq1},\ref{eq2}) contain the functional derivatives acting in the same
spatial point that demands a regularization. One of ways to remove
infinities from Eq. (\ref{eq2}) is a discretization, which,
for example, can be made by the means of a triangulation [4]. For our
case, it is sufficient to choose an elementary discretization by
introducing a cube spatial grid with the edge
length $\ell $. One could identify the scale of discretization with
the Planck length, however, it is not obligatory. As the  space is
divided into the cells with the volume $\Delta x\Delta y\Delta z =
\ell ^{3}$ and the centers located at points $\bm r_{1} ,\bm r_{2}
........\bm r_{k} $, it is necessary to replace a functional
derivative with usual derivative by a rule ${\textstyle{{\delta}
\over {\delta \xi ^{A}\left( {r} \right)}}} \to {\textstyle{{1}
\over {\ell ^{3}}}}{\textstyle{{\partial} \over {\partial \xi
_{k}^{A}} }}$, where it is implied that $\bm \xi _{k} $ is a value
of the vector $\bm \xi $ at the point $\bm r_{k} $, i.e. $\bm \xi
_{k} = \bm \xi \left( {\bm r_{k}} \right)$. As a result, Eq. (\ref{eq2}) will have the same form at all spatial points
$\bm r_{k} $ and its solution will be represented in the form of
product of solutions obtained for every spatial point:
\begin{equation}
\label{eq4} \Psi \left( {\bm \xi _{1} ,\bm \xi _{2} ...\bm \xi
_{k}} \right) = \psi \left( {\bm \xi _{1}}  \right)\psi \left(
{\bm \xi _{2}} \right) \ldots \psi \left( {\bm \xi _{k}} \right).
\nonumber
\end{equation}

The choice of operator ordering in the form of Laplacian leads to
the following equation
\begin{equation}
\label{eq5} \frac{{1}}{{\sqrt {G}} }\frac{{\partial} }{{\partial
\xi ^{A}}}\left( {\sqrt {G} \,G^{AB}\frac{{\partial} }{{\partial
\xi ^{B}}}} \right)\psi \left( {\bm \xi } \right) = 0,
\end{equation}
where $G = \det G_{AB} = 1/\det G^{AB}$. In Eq.
(\ref{eq5}) and everywhere further, the dependence on a spatial index
$k$ is omitted. One has to note, that for a case of pure
gravitation (i.e. in absence of of scalar fields), the equation
coinciding with (\ref{eq5}) can be written in the form of
\begin{equation}
\label{eq6} \gamma \,\mathop {\hat {\pi} }\nolimits^{ij} \left(
{\frac{{1}}{{\gamma }}\mathop {\tilde {G}}\nolimits_{ijkl}
\,\mathop {\hat {\pi} }\nolimits^{kl} } \right)\psi \left( {\gamma
_{mp}}  \right) = 0,
\end{equation}
where

\begin{equation}
\label{eq7} \mathop {\hat {\pi} }\nolimits^{ij} = \left\{
{{\begin{array}{*{20}c}
 {\frac{{\partial} }{{\partial \gamma _{ij}} },\;\;\;\;i = j,} \hfill \\
 {} \hfill \\
 {\frac{{1}}{{2}}\frac{{\partial} }{{\partial \gamma _{ij}} },\;\;\;\;i \ne
j.} \hfill \\
\end{array}} } \right.
\nonumber
\end{equation}

Let us find the solution of Eq. (\ref{eq5}) in the
form of ``plane waves '' [5,6,7]. Here, we introduce the following
variables $\tilde {u} = k^{ij}\gamma _{ji}$,  $\tilde {v} =
k^{ij}\gamma _{jm} k^{mn}\gamma _{nj}$ and $\Phi = p^{i}\phi _{i} $,
where $k^{ij}$ is some 3$\times$3-dimensional matrix and $p^{i}$ is
a vector of the dimension defined by a number of scalar fields $N$. Let us
represent the state $\psi$ in the form
\begin{equation}
\label{eq8} \psi \left( {\gamma _{lm} ,\phi _{i}}  \right) =
f\left( {\tilde {u},\tilde {v},\gamma}  \right)\exp\left( {i\Phi}
\right).
\end{equation}

Substitution of the expression (\ref{eq8}) into (\ref{eq5}) or, for pure
gravitation, into (\ref{eq6}) after cumbersome calculations using the Mathematica computer algebra
result in the following equation for the
function $f\left( {\tilde {u},\tilde {v},\gamma} \right)$:

\bea
 \frac{{6\gamma ^{ - 1/3}}}{{M_{p}^{2}} }
 \biggl(  - 3\gamma
^{2}\frac{{\partial ^{2}f}}{{\partial \gamma ^{2}}} - \biggl( {5 +
\frac{{N}}{{2}}} \biggr)\gamma \frac{{\partial f}}{{\partial
\gamma} } + 2\biggl( {\tilde {u}^{2} + \biggl( {\frac{{7}}{{3}} -
\frac{{N}}{{6}}} \biggr)\tilde {v}} \biggr)\frac{{\partial
f}}{{\partial \tilde {v}}}~~~~~~~~~~~~~~\nonumber\\ + 4\biggl( {2
\tilde {u}^{2}\tilde {v} - \tilde {u}^{4} + 8k\tilde {u}\gamma}
\biggr)\frac{{\partial ^{2}f}}{{\partial \mathop {\tilde
{v}}\nolimits^{2}} }  + \biggl( {\frac{{7}}{{3}} -
\frac{{N}}{{6}}} \biggr)\tilde {u}\frac{{\partial f}}{{\partial
\tilde {u}}} + \biggl( {2\tilde {v} - \tilde {u}^{2}}
\biggr)\frac{{\partial ^{2}f}}{{\partial \tilde {u}^{2}}}\nonumber
\\+ 4\biggl( {2\tilde {u}\tilde {v} - \tilde {u}^{3} + 6k\gamma}
\biggr)\frac{{\partial ^{2}f}}{{\partial \tilde {u}\partial \tilde
{v}}} - {2}\tilde {u}\gamma \frac{{\partial ^{2}f}}{{\partial
\tilde {u}\partial \gamma} } - 4\tilde {v}\gamma \frac{{\partial
^{2}f}}{{\partial \tilde {v}\partial \gamma} } \biggr) -
\frac{{1}}{{2}}p^{2}\gamma ^{ - 1/3}f = 0, \label{eqq}
 \eea
where $p^{2} = \left( {p^{1}} \right)^{2} + \left( {p^{2}}
\right)^{2} + \ldots \left( {p^{N}} \right)^{2}$ and $k = \det
k^{ij}$.

It is convenient to present a metric tensor in the form of $\gamma
_{ij} = a^{2}\mathop {\tilde {\gamma} }\nolimits_{ij} $, so that
$\det\mathop {\tilde {\gamma} }\nolimits_{ij} = 1$. The matrix
$\mathop {\tilde {\gamma }}\nolimits_{ij} $ describes a so-called
conformal geometry [8]. Then Eq. (\ref{eqq}) rewritten in
the terms of new variables $a = \gamma ^{1/6},\;u = k^{ij}\mathop
{\tilde {\gamma} }\nolimits_{ij} = \tilde {u}\gamma ^{ -
1/3},{\kern 1pt} \;v = k^{ij}\mathop {\tilde {\gamma}
}\nolimits_{jl} \,{\kern 1pt} k^{lm}\mathop {\tilde {\gamma}
}\nolimits_{mi} = \tilde {v}\gamma ^{ - 2/3}$ takes the form
\begin{equation}
\label{eq9}
\begin{array}{l}
 \frac{{1}}{{2M_{p}^{2}} }\left( { - \frac{{\partial ^{2}f}}{{\partial
a^{2}}} - \frac{{\left( {5 + N} \right)}}{{a}}\frac{{\partial
f}}{{\partial a}} + \frac{{8}}{{a^{2}}}\left( {5u\,\frac{{\partial
f}}{{\partial u}} + \left( {3v - u^{2}} \right)\frac{{\partial
^{2}f}}{{\partial u^{2}}} + \left( {3u^{2} + 11v}
\right)\frac{{\partial f}}{{\partial v}}} \right.}
\right. \\
 \left. {\left. { + 2\left( {24ku + v^{2} + 6u^{2}v - 3u^{4}}
\right)\frac{{\partial ^{2}f}}{{\partial v^{2}}} + 2\left( {18k +
7v u - 3u^{3}} \right)\frac{{\partial
^{2}f}}{{\partial u\partial v}}} \right)} \right)\\
~~~~~~~~~~~~~~~~~~~~~~~~~~~~~~~~~~~~~~~~~~~~~~~~~~
~~~~~~~~~~~~~~~~~~~~~~~~~~~~~~-\frac{{1}}{{2a^{2}}}p^{2}f = 0.
 \end{array}
 \nonumber
\end{equation}

A straightforward way is to solve the above
equation by the method of variable separation $f\left( {a,u,v}
\right) = R\left( {a} \right)g\left( {u,v} \right)$. Namely, if
the solution $g(u,v)$ of the equation
\begin{equation}
\label{eq10}
\begin{array}{l}
 5u\,\frac{{\partial g}}{{\partial u}} + \left( {3v - u^{2}}
\right)\frac{{\partial ^{2}g}}{{\partial u^{2}}} + \left( {3u^{2}
+ 11v} \right)\frac{{\partial g}}{{\partial v}} + 2\left(
{24k{\kern 1pt} u + v^{2}
+ 6u^{2}v - 3u^{4}} \right)\frac{{\partial ^{2}g}}{{\partial v^{2}}} \\
 \quad \quad \quad \quad \quad \quad \quad \quad \quad \quad \quad \quad
\quad \quad \quad + 2\left( {18k + 7v{\kern 1pt} u - 3u^{3}}
\right)\frac{{\partial ^{2}g}}{{\partial u\partial v}} = - \lambda g,\;\; \\
 \end{array}
\end{equation}
has obtained and the value of the corresponding constant $\lambda$
has been found, then the equation for the  function $R\left( {a}
\right)$ becomes
\begin{equation}
\label{eq11}
 - \frac{{\partial ^{2}R}}{{\partial a^{2}}} - \frac{{\left( {5 + N}
\right)}}{{a}}\frac{{\partial R}}{{\partial a}} - \frac{{8\lambda
+ M_{p}^{2} \,p^{2}}}{{a^{2}}}R = 0.
\end{equation}

The solution of (\ref{eq11}) can be expressed easy through
the Bessel functions. Thus, the main problem is to solve Eq. (\ref{eq10}), which describes quantization of the conformal
geometry [8] determined by a matrix $\mathop {\tilde {\gamma}
}\nolimits_{ij} $ with an unit determinant.

It is interesting to note that Eq. (\ref{eq10}) can be
obtained by another way.

Let us consider the  Hamiltonian
\begin{equation}
\label{eq12} H = \frac{{1}}{{2}}\left( {\mathop {\tilde {\gamma}
}\nolimits^{ij} }\right)'{\tilde {\gamma} }'_{ij}
 = \frac{{1}}{{2}}{\tilde {\gamma} }'_{ik} \,\mathop
{\tilde {\gamma} }\nolimits^{kl} \,{\tilde {\gamma} }'_{lj}
\,\mathop {\tilde {\gamma} }\nolimits^{ji} ,
\end{equation}

\noindent where $\det\mathop {\tilde {\gamma} }\nolimits_{ij} = 1$, and
a prime means a derivative on time. The Hamiltonian (\ref{eq12})
corresponds to free motion of a "particle" on a five-dimensional
surface $\det\mathop {\tilde {\gamma} }\nolimits_{ij} = 1$ of
constant negative curvature.

Firstly, let us  first consider  a simple case, when the matrix $\mathop
{\tilde {\gamma} }\nolimits_{ij} $ has dimension $2\times2$ and the
dimension of the surface $\det\mathop {\tilde {\gamma} }\nolimits_{ij} = 1$
equals two. After some parametrization $\mathop
{\tilde {\gamma }}\nolimits_{ij} \left( {\xi ^{1}\left( {t}
\right),\xi ^{2}\left( {t} \right)} \right)$, one has
\[
H = \frac{{1}}{{2}}G_{AB} \left( {\tau}  \right)\xi ^{A'}\left(
{t} \right)\xi ^{B'}\left( {\tau}  \right),
\]
where $G_{AB} = Tr\left[ {\tilde {\gamma} ^{ - 1}\frac{{\partial
\tilde {\gamma} }}{{\partial \xi ^{A}}}\tilde {\gamma} ^{ -
1}\frac{{\partial \tilde {\gamma} }}{{\partial \xi ^{B}}}}
\right]$ and a prime means differentiation on time $\tau $.

The generalized momentums is written in the form
\begin{equation}
\label{eq13} p_{A} = \frac{{\partial H}}{{\partial \xi ^{A\,'}}} =
G_{AB} \xi ^{B'\,}. \nonumber
\end{equation}

The Hamiltonian can be expressed through the momentums:
\begin{equation}
\label{eq14} H = \frac{{1}}{{2}}G^{AB}\left( {\tau}  \right)p_{A}
\left( {\tau} \right)p_{B} \left( {\tau}  \right). \nonumber
\end{equation}

Finally, quantization leads to the Schr\"{o}dinger equation
\[
\frac{{1}}{{\sqrt {G}} }\frac{{\partial} }{{\partial \xi
^{B}}}\left( {\sqrt {G} \,G^{AB}\frac{{\partial} }{{\partial \xi
^{A}}}\Theta}  \right) = \lambda \,\Theta .
\]

It is convenient to take the  coordinates
$\{ r\left( {\tau} \right),\varphi \left( {\tau} \right)\} $ for the parametrization $\{ \xi ^{1}\left(
{\tau} \right),\xi ^{2}\left( {\tau} \right)\} $  and
to represent a $2\times2$- matrix with the unit determinant as the product of three matrixes:
\bea \label{eq15} \mathop {\tilde {\gamma}
}\nolimits_{ij} = \left( {{\begin{array}{*{20}c}
 {\cos\varphi /2} \hfill & { - \sin\varphi /2} \hfill \\
 {\sin\varphi /2} \hfill & {\cos\varphi /2} \hfill \\
\end{array}} } \right)
\left( {{\begin{array}{*{20}c}
 {\exp r} \hfill & {0} \hfill \\
 {0} \hfill & {\exp\left( { - r} \right)} \hfill \\
\end{array}} } \right)~~~~~~~~~~~~~~~~~~~~~~~~~
\nonumber\\
\left( {{\begin{array}{*{20}c}
 {\cos\varphi /2} \hfill & {\sin\varphi /2} \hfill \\
 { - \sin\varphi /2} \hfill & {\cos\varphi /2} \hfill
\end{array}} } \right)
 = \left( {{\begin{array}{*{20}c}
 {\xi - \varsigma}  \hfill & {\sigma}  \hfill \\
 {\sigma}  \hfill & {\xi + \varsigma}  \hfill \\
\end{array}} } \right),\nonumber
\eea
where
\[
\varsigma = r \cos\varphi ,\, \sigma = r\sin\varphi , \,\xi = r.
\]
It turns out to be that the surface $\mathop
{\det\tilde {\gamma }}\nolimits_{ij} = 1$ for the $2\times2$- matrixes in coordinates
$r,\varphi $ is a hyperboloid $\xi ^{2} - \varsigma ^{2} - \sigma
^{2} = 1$. More precisely, the coordinates $r,\varphi $ parameterize
one of its cavities. Let us now search a solution of the
Schr\"{o}dinger equation in the form of $\Theta \left( {r,\varphi}
\right) = g\left( {k^{ij}\tilde {\gamma} _{ij}} \right)$. That leads to the following equation
\[
 - \frac{{1}}{{2}}u^{2}{g}''\left( {u} \right) - u{g}'\left( {u} \right) =
\lambda g\left( {u} \right),
\]
where $u = k^{ij}\tilde {\gamma} _{ij} $. A solution of the above
equation can be written as
\begin{equation}
\label{eq16} g\left( {u} \right) = u^{ - 1/2 + i\sqrt {2{\kern
1pt} \lambda - 1/4}} .
\end{equation}

Using the matrix $k^{ij}$ in the form of
\begin{equation}
\label{eq17} k^{ij} = \left( {{\begin{array}{*{20}c}
 {\cos\theta {\kern 1pt} /2} \hfill & { - \sin\theta {\kern 1pt} /2} \hfill \\
 {\sin\theta {\kern 1pt} /2} \hfill & {\cos\theta {\kern 1pt} /2} \hfill \\
\end{array}} } \right)\left( {{\begin{array}{*{20}c}
 {1} \hfill & {0} \hfill \\
 {0} \hfill & {0} \hfill \\
\end{array}} } \right)\left( {{\begin{array}{*{20}c}
 {\cos\theta {\kern 1pt} /2} \hfill & {\sin\theta {\kern 1pt} /2} \hfill \\
 { - \sin\theta {\kern 1pt} /2} \hfill & {\cos\theta {\kern 1pt} /2} \hfill \\
\end{array}} } \right),
\nonumber
\end{equation}
gives
\[
u = k^{ij}\mathop {\tilde {\gamma} }\nolimits_{ij} = \xi - n_{1}
\varsigma - n_{2} \sigma ,
\]
where $n_{1} = \cos\theta ,{\kern 1pt} \,n_{2} = \sin\theta $. Thus, (\ref{eq16}) is precisely a plane wave [5,6,7] on a
surface of hyperboloid in a $2+1$-dimensional Minkowski space.

For the case of the $3\times3$-dimensional matrixes, we will search a solution in the form of $\Theta \left( {\xi ^{1},\xi ^{2},\xi
^{3},\xi ^{4},\xi ^{5}} \right) = g\left( {u,v} \right)$, where $u
= k^{ij}\mathop {\tilde {\gamma} }\nolimits_{ij},\, v =
k^{ij}\mathop {\tilde {\gamma} }\nolimits_{jl} \,{\kern 1pt}
k^{lm}\mathop {\tilde {\gamma} }\nolimits_{mi} $, and  $k^{ij}$ is
some matrix. After the cumbersome calculations, we come again to Eq. (\ref{eq10}). Since the wave function has five
independent coordinates, the function $\Theta $ should be
defined by the value of the constant $\lambda $ and some additional four
parameters. Thus, it is possible to impose at least two additional
conditions on the tensor $k^{ij}$.   It should be  noted, that
Eq. (\ref{eq10}) becomes
homogeneous relatively $v$ and $u^{2}$ when $k = \det k^{ij} = 0$. This allows finding
the solution of (\ref{eq10}) at $k = 0$ in the form of
\begin{equation}
\label{eq18} g\left( {u,v} \right) = \left( {u^{2} - v}
\right)^{i\alpha /2 - 3/4}s\left( {u^{2}/v} \right).
\end{equation}

Substitution of (\ref{eq18}) into (\ref{eq10}) results in the equation
for the function $s\left( {z} \right)$

\bea
\label{eq19} 3\left( {z - 1} \right)\left( {2\left( {z - 2}
\right)\left( {z - 1} \right)z{s}''\left( {z} \right) + \left(
{z\left( {4z - 7} \right) + 2} \right){s}'\left( {z} \right)}
\right)\nonumber\\ - \left( {\lambda - \frac{{\alpha ^{2}}}{{2}} -
\frac{{9}}{{8}}} \right)s\left( {z} \right) = 0,
\eea
with the general solution expressed through the hypergeometric
function
\bea
\label{eq20}
\begin{array}{l}
 s( {z} ) =
 \frac{{c_{1}} }{{\sqrt {2 - {\textstyle{{2} \over
{z}}}}} }\,_{2}F_{1} \biggl( {\textstyle{{1} \over {12}}}\bigl( {9
- 2\sqrt {3} \,i\sqrt {2\lambda - 3 - \alpha ^{2}}}
\bigr),~~~~~~~~~~~~~~~~~~~~~~~~~~~\\~~~~~~~~~~~~~~~~~
{\textstyle{{1} \over {12}}}\bigl( {2\sqrt {3\,} i\sqrt {2\lambda
- 3 - \alpha ^{2}} + 9} \bigr);{\textstyle{{3} \over
{2}}};{\textstyle{{z} \over {2( {z - 1})}}}\biggr)
~~~~~~~~~~~~~~~~~~~~\\~~~~~~~~~~~~~~~~~~~~~+
 c_{2} \,_{2} F_{1} \biggl( \frac{{1}}{{12}}\bigl( {3 - 2\sqrt {3}
\,i\sqrt {2\lambda - 3 - \alpha ^{2}}}
\bigr)\\~~~~~~~~~~~~~~~~~~~~~~~~~~~~~~~~,\frac{{1}}{{12}}\bigl(
{2\sqrt {3} \,i\sqrt {2\lambda - 3 - \alpha ^{2}} + 3}
\bigr);\frac{{1}}{{2}};\frac{{z}}{{2( {z - 1} )}} \biggr).
 \end{array}
\eea

Thus, the wave function $\Theta $ depends on parameters $\alpha $,
$\lambda $ and some additional three parameters contained in the tensor
$k^{ij}$, which obeys the condition $k = 0$ and some two additional
conditions.

Here, we do not discuss a question about normalization of the wave
function in detail. The detailed consideration of quantization on a hyperboloid is presented in Ref. [9]. Here, we
note that for the wave function to be normalized, it is
necessary, that the normalization of wave function requires positivity of the expression $2\lambda - 3 - \alpha ^{2}$ under
the square root in Eq. (\ref{eq20}).
The minimal possible value of the constant $\lambda = 3/2$ is reached
at $\alpha = 0$.

In the given work we have solved the discrete Wheeler-DeWitt equation in the vicinity of small
scale factors by the method of the variables
separation. It is shown, that the constant $\lambda $ of the variables
separation cannot be infinitely small. The minimal
admissible value of $\lambda $ arises because the quantization is carried out on
a surface of constant negative curvature.

It means that, if the scale $\ell $ of
discretization is of order of the inverse Planck mass, there exists an energy density $\rho =
\frac{{8}}{{M_{p}^{2} \ell ^{6}a^{6}}}\lambda \approx
\frac{{12M_{p}^{4}} }{{a^{6}}}$ in early Universe due to fluctuations of
the conformal geometry. Let us emphasize that there is a time, when the matter fields as well as the gravitational waves do not oscillate
so that the well-known vacuum energy corresponding to zero
oscillations of the field oscillators is not formed
yet.

\acknowledgement{The authors are grateful to  Victor Red'kov for
the discussion.}

\newpage

\noindent REFERENCES

\par\noindent  [1] J.A. Wheeler  Superspace and Nature of Quantum
Geometrodynamics. In: DeWitt C., Wheeler J.A. (eds.) Battelle
Rencontres, (New York: Benjamin, 1968).
\par\noindent
[2] B.S. DeWitt, Phys. Rev. 160, 1113 (1967).
\par\noindent
[3] S. L. Cherkas, V. L. Kalashnikov, Gen. Rel. Grav.  44, 3081
(2012).
\par\noindent
[4] H. W. Humber, Gen. Rel. Grav.  41, 817 (2009).
\par\noindent
[5] I.M. Gelfand, M.I. Graev, N.J. Vilenkin Integral geometry and
the related questions of the theory of representations. (Moscow:
State Edition of Physics and Mathematics literature, 1962) [in
Russian].
\par\noindent
[6] I.S. Shapiro, Dokl. Akad. Nauk SSSR, 106, 647 (1956).
\par\noindent
[7] E. M. Ovsiyuk, N. G. Tokarevskaya, V. M. Red'kov,  Nonlin.
Phenomena Complex Syst.  12, 1 (2009).
\par\noindent
[8] J. W. Jr. York,   Phys. Rev. Lett. 26, 1656 (1971).
\par\noindent
[9] V.S. Otchik , V.M. Red'kov, Preprint of Institute of Physics
Akad. Nauk BSSR No. 298 (Minsk: Institute of Physics, 1983).

\end{document}